\documentclass[proof]{pasj00}
\draft

\begin{document}
\SetRunningHead{Author(s) in page-head}{Running Head}

\title{Near- and Mid-infrared Imaging Study \\ of Young Stellar Objects around LkH$\alpha$ 234\thanks{Based in part on data collected at Subaru Telescope, which is operated by the National Astronomical Observatory of Japan.}}

\author{Eri \textsc{Kato},\altaffilmark{1}
            Misato \textsc{Fukagawa},\altaffilmark{1}
            Marshall \textsc{D. Perrin},\altaffilmark{2}  
            Hiroshi \textsc{Shibai},\altaffilmark{1}
            Yusuke \textsc{Itoh},\altaffilmark{1}
            Takafumi \textsc{Ootsubo},\altaffilmark{3}
            }
\altaffiltext{1} {Department of Earth and Space Science, Graduate School of Science, Osaka University,\\ 1-1 Machikaneyama, Toyonaka, Osaka 560-0043, Japan}
\altaffiltext{2} {UCLA/NSF Center for Adaptive Optics}
\altaffiltext{3} {Astronomical Institute, Tohoku University,\\ 6-3 Aramaki, Aoba-ku, Sendai, Miyagi 980-8578, Japan}


%

\maketitle

\begin{abstract}
We present high-resolution ($0\farcs2$) near-infrared images of the area surrounding the Herbig Be star LkH$\alpha$~234 taken with the Coronagraphic Imager with Adaptive Optics (CIAO) and the adaptive optics on the Subaru Telescope. The near-infrared ($J$, $H$, $K$, $L'$ and $M'$ bands) images reveal circumstellar structures around LkH$\alpha$~234 in detail. Eight young stellar object (YSO) candidates (Object~B, C, D, E, F, G, NW1, and NW2) were detected at 2$\arcsec$--11$\arcsec$ from LkH$\alpha$~234. Objects~B and C are likely variable stars, which is consistent with a young evolutionary status. Three objects (LkH$\alpha$~234, NW1, and NW2) were identified in the 11.8~$\micron$ (SiC) and 17.65~$\micron$ images obtained with the Long Wavelength Spectrometer (LWS) on the Keck Telescope. NW1 and NW2 are thought to be embedded young stars. We suggest that NW1 and not LkH$\alpha$~234 is the source illuminating the reflection nebula west of LkH$\alpha$~234, although Object~G may be another candidate. In our images, these objects are located at the center of the 2~$\micron$ polarization, and NW1 resides at the center of a monopolar cavity. The SED of NW1 suggests that it is a YSO with a spectral type of B6--7. Object~F and G were discovered by our observations at $1\farcs9$ and $2\farcs3$, respectively, from LkH$\alpha$~234, and their proximity to LkH$\alpha$~234 suggests that they may be its lower-mass companions.
 \end{abstract}

\section{Introduction}

To understand star and planet formation mechanisms, it is necessary to understand the properties of circumstellar environments such as cluster density, circumstellar disks and envelopes. Although many stars are formed within clusters (\cite{LadaLada2003}), the process of cluster formation is not clearly understood. Other unanswered questions remain; for example, what is the difference between the formation of single and binary components? Are companion stars formed via fragmentation of a molecular core or disks (\cite{Wheelwright2010}, \cite{Baines2006})?
Moreover, the nature of protoplanetary disks affects the process of planet formation. Past observations at radio and infrared (IR) wavelengths show that circumstellar disks are ubiquitous around young stars (e.g., \cite{Andrew2006}). Observations at high-angular resolutions have frequently revealed a complex distribution of circumstellar material, which might be attributed to stellar multiplicity and properties of the local star-forming environments such as initial disk masses (e.g., \cite{Fukagawa2010}). Therefore, it is important to investigate the environments of young stellar objects (YSOs) in the phase of star formation on a small spatial scale.

Herbig Ae/Be (2~${M}_{\odot }$ $\lesssim$ ${M}_{\star}$ $\lesssim$ 10~${M}_{\odot }$) stars are intermediate-mass pre-main-sequence stars that are in evolutionary stages similar to lower mass T Tauri (0.1~${M}_{\odot}$ $\lesssim$ ${M}_{\star}$ $\lesssim$ 2~${M}_{\odot }$) stars. It is thought that the formation mechanism of low-mass stars is by the gravitational collapse of a single cloud core (\cite{Palla1993}). In contrast, the formation and evolution mechanisms of massive stars (${M}_{\star}$ $>$ 10~${M}_{\odot}$) are still under debate (\cite{Palla1993}, \cite{Wang2007}, \cite{Zinnecker2007}). Because they reside, at the boundary between low- and high-mass stars, Herbig Ae/Be stars can provide significant insight into star formation. In addition, they can be observed more easily than massive stars because their environments are comparatively simple (\cite{Testi1999}, \cite{Wang2007}, \cite{LadaLada2003}, \cite{Hillenbrand1997}). Further, because Herbig Ae/Be stars are more common than massive stars, a large number of stars including much closer ones can be investigated. However, the detailed circumstellar environments around Herbig Ae/Be stars, especially Herbig Be stars, remain less understood than those around T Tauri stars. This is partly because it is difficult to detect such stars owing to their rapid evolution. Moreover, because the youngest Herbig Ae/Be stars are typically located at large distances ($\sim$kpc), high-spatial resolution is key to understanding their nature.    



The Herbig Be star, LkH$\alpha$~234, is located in the NGC 7129 star-forming region at 1.25~kpc (\cite{Schevchenko1989}). 
LkH$\alpha$~234 is a star of spectral type B5$\sim$7, mass~8.5~$M_{\odot}$, luminosity~$\sim$1700 $L_{\odot }$, and estimated age~$\sim$10$^5$~yr (\cite{Hillenbrand1992}, \cite{Fuente2001}). The NGC~7129 star-forming region was previously observed at various wavelengths from the optical to radio. Another deeply embedded YSO located at 3$\arcsec$ from LkH$\alpha$~234 was detected by 2~$\micron$ polarization observations; this could be a protostar that illuminates the reflection nebula (\cite{Wein1994}, \cite{Wein1996}). Millimeter-wave interferometric observations reveal two objects (IRS~6 and MM1) at a few arcseconds from LkH$\alpha$~234, and a large CO outflow is observed near them (\cite{Fuente2001}). The position of the protostar inferred by the 2~$\micron$ polarimetry is different from that inferred from the millimeter-wave interferometric observations. 
Other clumps (at $4\farcs4$, $3\farcs0$, $2\farcs1$ and $2\farcs2$ from LkH$\alpha$~234) and H$_{2}$O maser spots were detected by observations of the radio continuum and the H$_{2}$O maser line with the Very Large Array (VLA; \cite{Trini2004}, \cite{Ume2002}). These previous observations showed that the CO outflow detected around this region is driven not by LkH$\alpha$~234 but by other objects. In addition, H$_2$ and [S~II] jets observed in this region do not point toward LkH$\alpha$~234, which suggests that there are other unseen sources producing the jets (\cite{Ray1990}, \cite{Cabrit1997}). To understand the unresolved sources driving these complicated outflows, it is essential to investigate the masses and locations of objects around LkH$\alpha$~234.

In this paper, we report observations of the Herbig Be star LkH$\alpha$~234 using the stellar coronagraphic camera CIAO on the Subaru Telescope and the LWS on the Keck Telescope. Our high-resolution images reveal detailed circumstellar~structures around LkH$\alpha$~234 and eight YSOs at 2$\arcsec$--11$\arcsec$ from LkH$\alpha$~234.

\section{Observations} 

\subsection{Near-infrared Observations}

LkH$\alpha$~234 was observed using the adaptive optics (AO) with Subaru/CIAO (\cite{Tamura2000}, \cite{Murakawa2004}). Near-infrared images were obtained in the $J$ (1.25 $\micron$), $H$ (1.64 $\micron$), and $K$ (2.20 $\micron$) bands on July 15, 2006. In addition, the images were taken in the $J$ band alone on May 25, 2008, and the $L'$ (3.77 $\micron$) and $M'$ (4.68 $\micron$) bands on July 10, 2008. The sky was clear all three nights. CIAO utilizes a 1024~$\times$~1024 InSb ALADDIN II array. The pixel scale of the medium-resolution mode was $0\farcs0213$ pixel$^{-1}$, and the field of view was $21\farcs8$~$\times$~$21\farcs8$. The observations (bands, exposure times, number of coadds, size of the occulting masks, photometric standard stars, and natural seeing) are summarized in Table \ref{tab:table_log}.
 
Adaptive optics (AO) allowed us to achieve high-spatial resolution given by the full-width at half-maximum (FWHM). On July 15, 2006, a spatial resolution was $0\farcs15$--$0\farcs2$ in the $J$, $H$, and $K$ bands, and on May 25, 2008, a spatial resolution of $0\farcs17$ was achieved in the $J$ band. Finally, on July 10, 2008, spatial resolutions of $0\farcs22$ and $0\farcs20$ were achieved in the $L'$ and $M'$ bands, respectively.


 

\subsection{Mid-infrared Observations}

Mid-infrared observations were obtained with the LWS (\cite{Jones1993}) on the W.~M.~Keck I Telescope. We obtained images using the 11.8~$\micron$ ``SiC'' filter ($\Delta \lambda = 2.32~\micron$) on July 26, 2004 and using the 17.65~$\micron$ filter ($\Delta \lambda = 0.85~\micron$) on August 28, 2004. Conditions were clear and dry on both nights. As measured from the Caltech Submillimeter Observatory (CSO) radiometer, the precipitable water vapor was 2.2~mm during the 11.8~$\micron$ SiC observations and only 1~mm during the 17.65 $\micron$ observations, thereby providing good atmospheric transmissivity. The LWS was used in the chop-nod imaging mode with a 10$\arcsec$ throw at a position angle of 0~degrees and a chop frequency of 2.5 Hz. The total on-source integration time was 285~s for observations with each filter. The point-spread function (PSF) FWHMs were $0\farcs32$ at 11.8~$\micron$ and $0\farcs47$ at 17.65~$\micron$. For photometric calibration, we observed HD~197989, a mid-infrared photometric standard, from the catalog of \citet{Cohen1999} in a similar manner. The observations (bands, exposure times, photometric standard stars, and natural seeing) are summarized in Table \ref{tab:table_log}.

\section{Data Reduction}

\subsection{Near-infrared Data Reduction}

The images were processed using the Image Reduction and Analysis Facility (IRAF) packages (\cite{Tody1986}) \footnote{IRAF is distributed by the National Optical Astronomy Observatories, which are operated by the Association of Universities for Research in Astronomy, Inc., under cooperative agreement with the National Science Foundation.} for dark subtraction, flat-fielding with sky flats, sky subtraction, and bad pixel correction. For the $L'$ and $M'$-band images, sky emission was cancelled by subtracting each pair obtained, by nodding the telescope in the north-south direction by 10$\arcsec$. The total exposure time was just half of the total time given above for Objects B, D and E (Section~4 for designations of objects), each of which were covered by only one pointing of the nodding pair.  

\begin{figure}
  \begin{center}
    \FigureFile(160mm,140mm){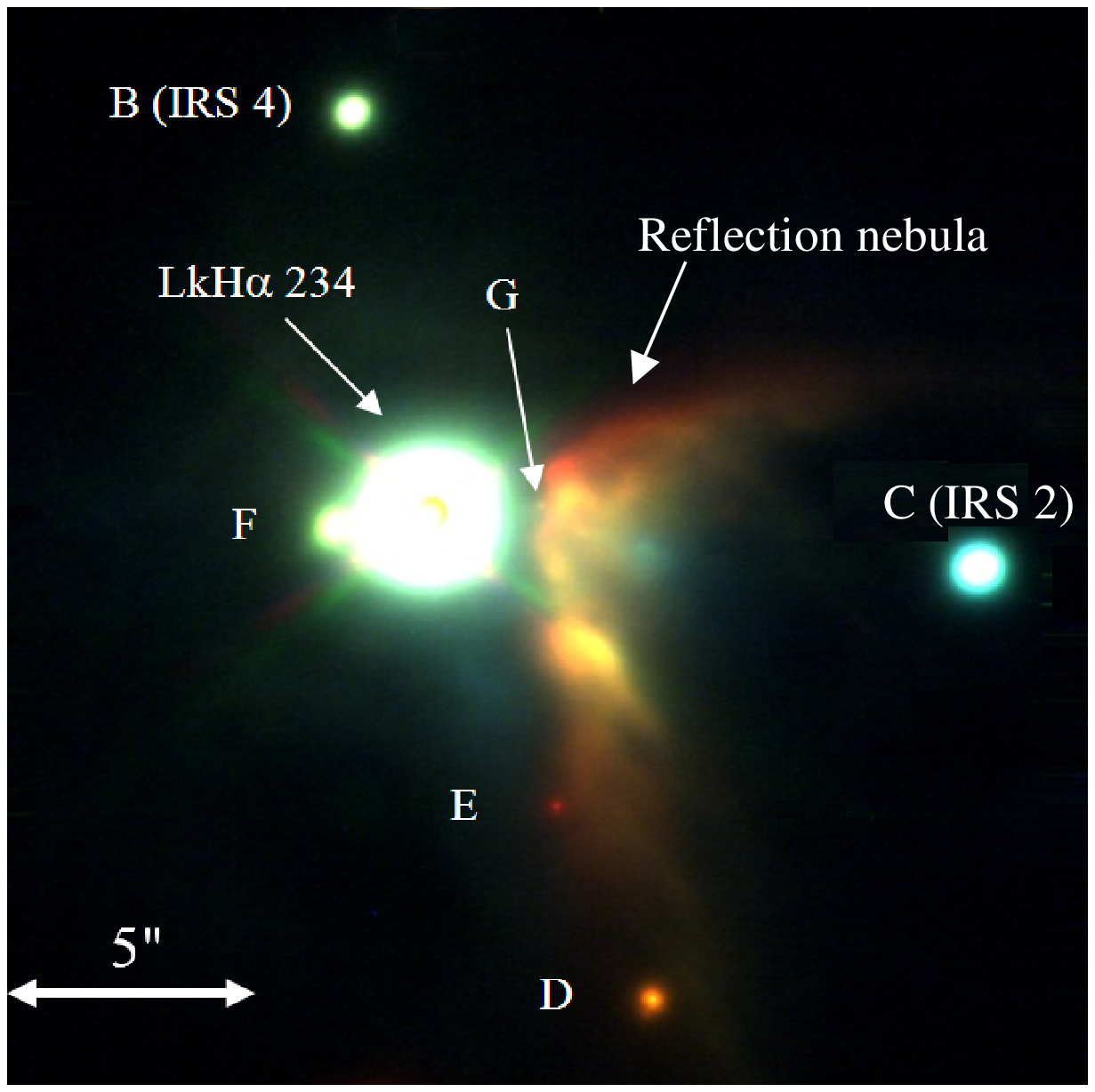}
  \end{center}
  \caption{$JHK$ color composite image around LkH$\alpha$~234 displayed in linear scale. Ghost images due to the beam splitter are software masked. North is up and east is to the left. }\label{fig:fig_1}
\end{figure}

Nine objects were detected in our near-infrared images in the field of view of $21\farcs8$~$\times$~$21\farcs8$ (LkH$\alpha$~234, Objects~B--G, and NW1--2; Figures \ref{fig:fig_1} and \ref{fig:fig_2}). We conducted PSF fitting and aperture photometry using the DAOPHOT and APPHOT IRAF packages. PSF fitting photometry was applied for Objects~F and G in the $JHK$ bands with an empirical PSF template made from Objects~B and C, because the magnitudes for these objects could not be accurately measured owing to contamination from the skirts of the PSF of LkH$\alpha$~234. Object~B, C, D, and F were measured by aperture photometry. The photometric values were corrected using the measurements of a photometric standard star with the same aperture (of about four times the FWHM). Relative photometry to Object~B was also performed with the same aperture as Object~B, which produced results consistent with the PSF fitting and aperture photometry results. Photometry in the $L'$ and $M'$ bands was performed by aperture photometry.

\subsection{Mid-Infrared Data Reduction}

The LWS data were reduced using a custom Interactive Data Language (IDL) pipeline based on the LWSCOADD routines made available by Keck Observatory plus extensions for atmospheric transmission correction, photometric calibration, and source detection. First, each chop-nod set was reduced by standard double-differencing to remove background emission. No flat fielding was performed. Bad pixels were identified and replaced by the median of neighboring pixels. The data were then rotated using a cubic spline interpolation to orient north up. All individual chop-nod-subtracted images were then registered via Fourier cross-correlation and summed into a mosaic image. 

To estimate a correction for atmospheric transmission losses, we first calculated a grid of Mauna Kea atmosphere models within a range of airmasses and water vapor contents, using the BFATS atmospheric transmission code (\cite{Roe2002}). For each model, we computed an overall atmospheric transmission weighted by the relevant transmission profile for each filter. Then, for each observation, given the measured water vapor (from the CSO radiometer) and airmass, we obtained a transmission correction by linearly interpolating between the closest models in the precomputed atmosphere grid.

Observations of standard stars were reduced in the same way as the science data. We followed the method provided by \citet{Cohen1999} for flux calibration based on determination of the instrumental relative spectral response (RSR) consisting of the product of transmission curves for the atmosphere, instrumental optics, filter, and detector quantum efficiency. We also derived aperture correction factors based on the radial profile of the standard star's PSF. Finally, we evaluated the absolute precision of these flux calibrations by comparing with the results of other authors; LWS photometry was calibrated using a near-identical procedure by \citet{McCabe2006}, and the agreement in absolute flux calibrations is within a few percent.

\section{Results \& Discussions}

\subsection{Nebulae Around LkH$\alpha$ 234}

Our imaging observations reveal remarkable nebulae around LkH$\alpha$~234. One of them, located at approximately $3\arcsec$ ($\sim$4000~AU) west of LkH$\alpha$~234, extends more than 10$\arcsec$ westward. In addition to appearing in the $J$, $H$,  and $K$ bands, it appears in the $L'$ and $M'$~bands (Figures \ref{fig:fig_1} and \ref{fig:fig_2}). Our high-spatial resolution images achieved with AO reveal its complex morphology. The nebula has three bright clump-like structures at the following distances and position angles (P.A.) from LkH$\alpha$~234: $2\farcs78$ $\pm $ $0\farcs3$ and \timeform{276D} $\pm $ \timeform{4D}, $2\farcs78$~$\pm$~$0\farcs25$ and \timeform{248D} $\pm $ \timeform{1D}, and $4\farcs22$ $\pm$ $0\farcs4$ and \timeform{231D} $\pm$ \timeform{1D}. The nebula has the shape of a monopolar cavity, suggestive of the outflow activity frequently seen around Class~I stars. A protostar might therefore be the source of the outflow. Infrared emission is often attributed to dust grains heated by a nearby source, which in this case may be an embedded protostar. However, the observed nebula extends more than 10$\arcsec$. In addition, near-infrared polarization observations of the nebula showed that it had a typical concentric pattern explained by scattered light (\cite{Wein1994}), indicating that it is a reflection nebula. The center of the polarization pattern does not coincide with LkH$\alpha$~234, which also suggests that the nebula is not illuminated by LkH$\alpha$~234, but by another object that is probably the source of the outflow. 

\begin{figure}
  \begin{center}
    \FigureFile(140mm,100mm){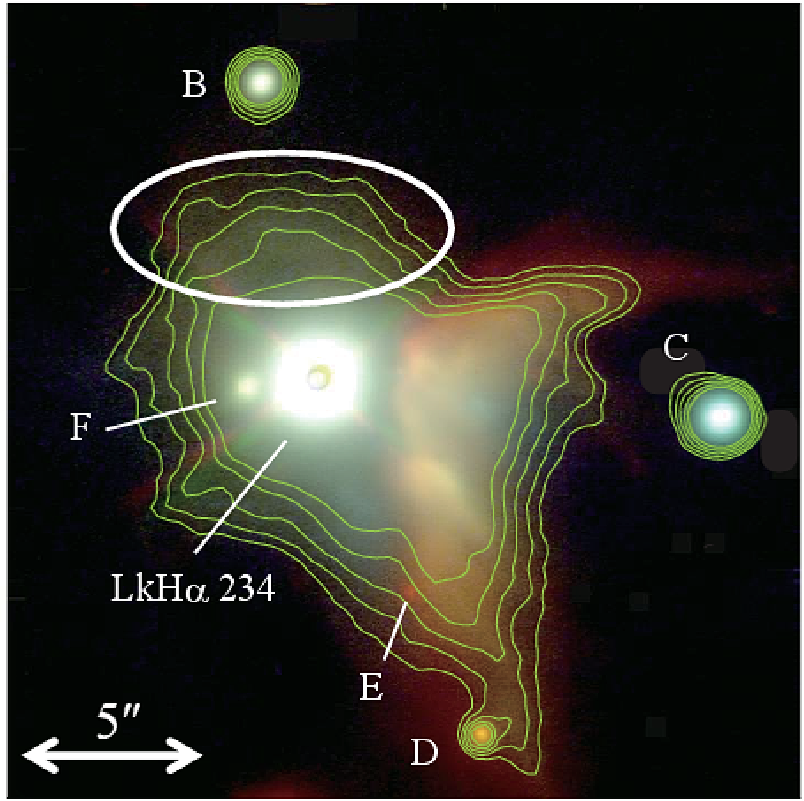}
  \end{center}
  \caption{$JHK$ image displayed in logarithmic scale to show the faint northeast nebula (indicated by the ellipse). The faint nebula does not appear in Figure \ref{fig:fig_1} because of the different display scale. The contour level is 0.7~$\sigma$ to 2.7~$\sigma$ in the $H$-band image. Ghost images, like the one in northeast of Object~C, are masked in this figure. Note the half-round structure of this nebula at approximately 5$\arcsec$ ($\sim$6000~AU) northeast of LkH$\alpha$~234. North is up and east is to the left. }\label{fig:fig_4}
\end{figure}

LkH$\alpha$~234 is surrounded by another tenuous nebula that appears green around it in Figure \ref{fig:fig_4}. This nebula extends approximately 5$\arcsec$ ($\sim$6000 AU) toward the northeast from LkH$\alpha$~234. No luminous stars can be seen except for LkH$\alpha$~234 near this northeast~nebula, suggesting that LkH$\alpha$~234 probably illuminates it. Imaging polarimetry of this northeast nebula using the Lick~AO system indicates its polarization position angle is consistent with its being scattered light from LkH$\alpha$~234 (\cite{Perrin2006}). The $J-H$ and $H-K$ colors of this northeast nebula measured at an offset ($\alpha$, $\delta$) = ($-0\farcs43$, $+3\farcs4$) from LkH$\alpha$~234 with an aperture radius of 20~pixels are 0.88 and 1.06~magnitudes, respectively. Meanwhile, the colors of LkH$\alpha$~234 are 1.33 magnitudes and 1.12 magnitudes. Because the color of this nebula is more blue than that of LkH$\alpha$~234, the scatterers of the starlight are not large grains (0.25--0.8~$\micron$) but probably smaller interstellar dust grains ($<$~0.25~$\micron$; \cite{Pend1990}).

\begin{figure}
  \begin{center}
    \FigureFile(160mm,160mm){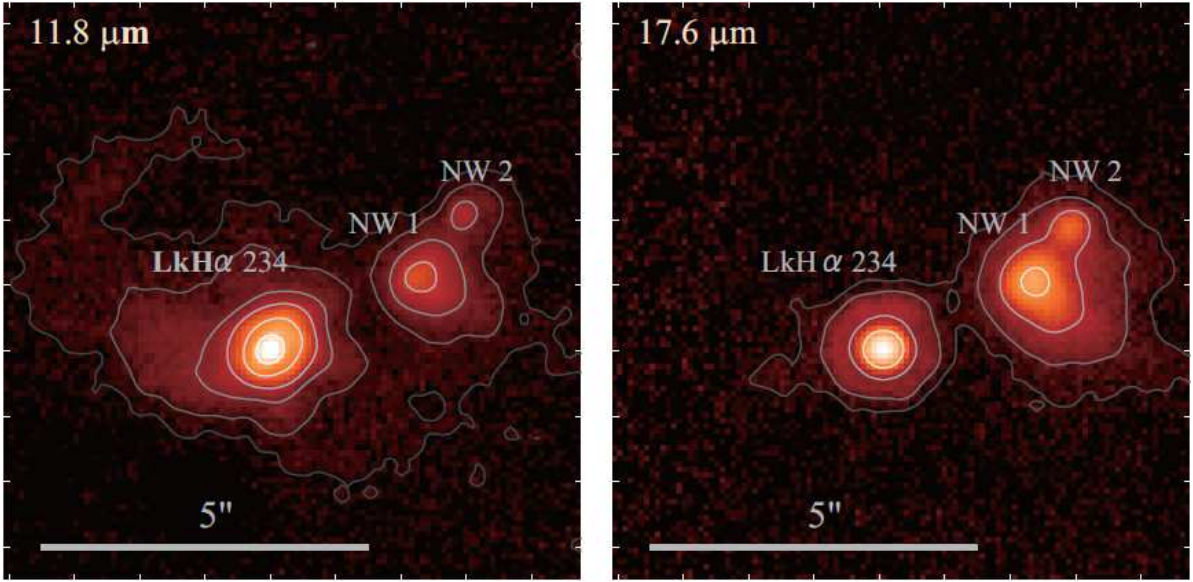}
  \end{center}
  \caption{(a) 11.8~$\micron$ (SiC; left panel) and (b) 17.65~$\micron$ (right panel) Keck LWS images around LkH$\alpha$~234 displayed in logarithmic scale. North is up and east is to the left. The field of view is 9$\arcsec$ $\times$ 9$\arcsec$ for both. Contours are shown logarithmically spaced from 0.001 to 0.33 of the peak intensity for the 11.8~$\micron$ image and from 0.01 to 0.33 of the peak intensity for the 17.65~$\micron$ image. Note that the embedded sources of NW1 and NW2 are bright at the mid-infrared wavelengths. In particular, NW1 is comparable in flux to LkH$\alpha$~234 at 17.65~$\micron$.  }\label{fig:fig_5}
\end{figure}

Figure \ref{fig:fig_5} shows the 11.8~$\micron$ (SiC) and 17.65~$\micron$ images. These images show three discrete objects (LkH$\alpha$~234, NW1, and NW2). NW1 and NW2 are located at $2\farcs52$ and $3\farcs69$, respectively, from LkH$\alpha$~234. These two objects are not seen in the $J$, $H$, and $K$-band images; however, they are bright in the mid-infrared bands, suggesting that NW1 and NW2 are embedded young stars. In contrast, Objects~F and G seen in the near-infrared images are not detected in the mid-infrared images, implying that they are relatively mature stars with less infrared excesses. Moreover, the 11.8~$\micron$ image shows a diffuse envelope-like nebula east of LkH$\alpha$~234 that curves northward of it. 
This is considered to be the same nebula as the tenuous northeast nebula discussed above. The region of diffuse 11.8~$\micron$ emission in Figure \ref{fig:fig_5} corresponds to the darker region just inside the northeast nebula's half-round structure in Figure \ref{fig:fig_4}. The curved mid-infrared nebula is probably a denser part of this northeast nebula. The mid-infrared nebular area just has a higher extinction; thus, it is opaque at less than 2~$\micron$, but it emits at longer wavelengths.

\subsection{YSO Candidates}

Eight YSO candidates (Objects~B, C, D, E, F, G, NW1, and NW2) were detected at $2\arcsec$--11$\arcsec$ from LkH$\alpha$~234 in the near- and mid-infrared images. Table~\ref{tab:table_1} lists the positions and photometric magnitudes of these objects. Table \ref{tab:table_4} lists objects observed by other authors that correspond to these objects. Four of them (Object~D, E, F, and G) were detected in the $K$ band for the first time. In particular, Objects~F and G were discovered by our observations. Objects~B and C correspond to IRS~4 and IRS~2, respectively named by \citet{Wein1994}; NW1 corresponds to NW detected by \citet{Polomski2002} and IRS~6 detected by \citet{Cabrit1997}. NW2 was detected at 11.8 and 17.65~$\micron$ as well as in the $L'$ and $M'$ bands. In total, nine stars of intermediate- and low-masses (LkH$\alpha$~234 and Objects~B, C, D, E, F, G, NW1, and NW2) were detected within 11$\arcsec$ ($\sim$14000~AU) of LkH$\alpha$~234, possibly forming a cluster. Below, we discuss the characteristics of the detected YSO candidates based on their positions in color-color diagrams, and we compare the present observational results with previous studies. 

\subsubsection{Color-color Diagrams}
\begin{figure}
  \begin{center}
    \FigureFile(170mm,100mm){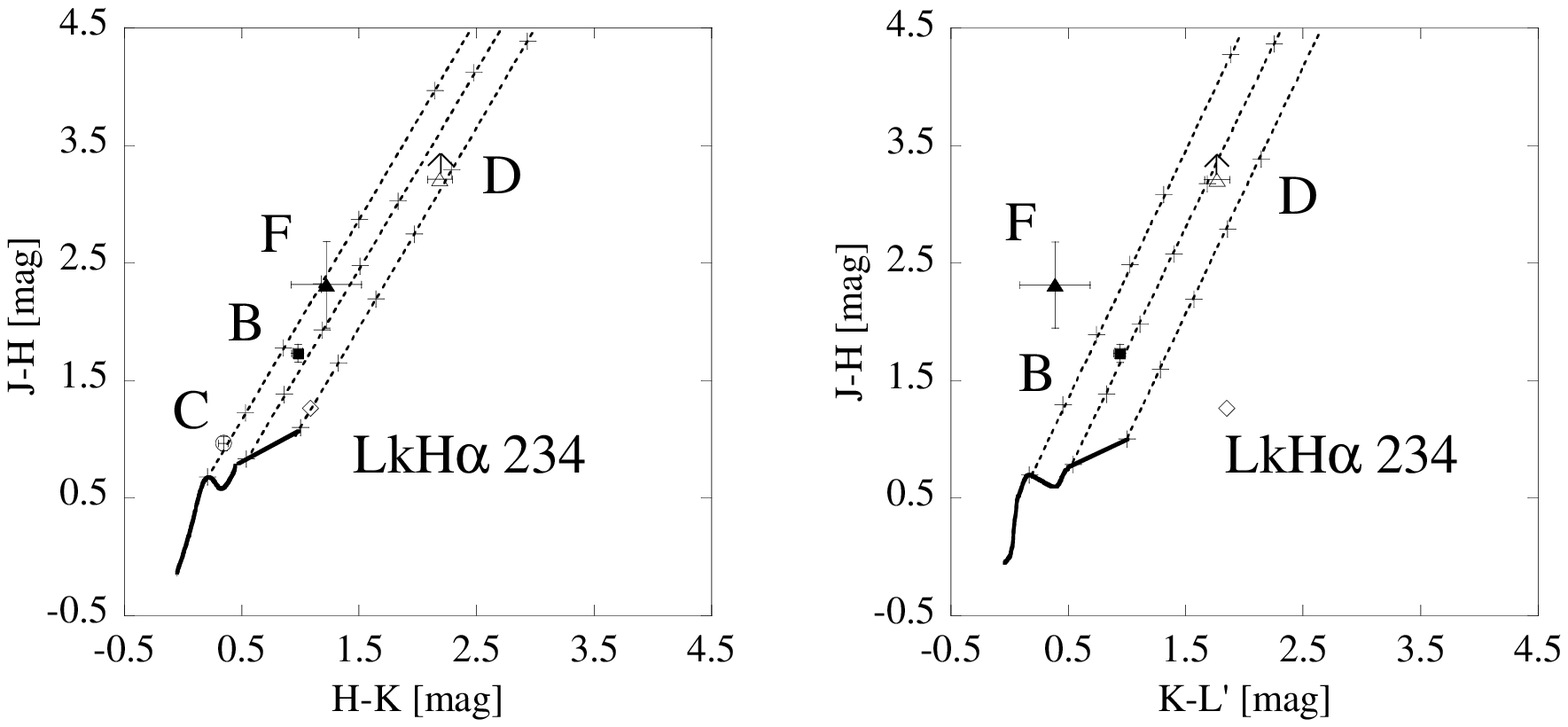}
  \end{center}
  \caption{(a) $JHK$ color-color diagram (left panel). (b) $JHKL'$ color-color diagram (right panel). Five objects (Objects~B, C, D, F and LkH$\alpha$~234) are plotted. The $JHK$ colors are transformed to the CIT photometric system (\cite{Cutri2003}, \cite{Itoh2001}). The bold curve shows colors of unreddened main-sequence (\cite{Lada1992} and \cite{Koorn1983}) and T Tauri (\cite{Mayer1997}) stars. The intrinsic colors of the main sequence were determined from the values for $V-K$ as a function of the spectral type (\cite{Koorn1983}), and spectral types from O6 to M8 were included. The three dotted lines are parallel to the reddening vector. In the $JHK$ color-color diagrams, the gradient of the reddening vector is 1.7 (\cite{Koorn1983}); in the $JHKL'$ color-color diagram it is 2.09 assuming the reddening law of \citet{Bessel1988} and \citet{Martin1990}. Crosses represent the amounts of extinction ($A_V$), which are 0, 5, 10, 15, 20, 30 and 40~magnitudes.  }\label{fig:fig_7}
\end{figure}

Figure \ref{fig:fig_7} shows $JHK$ and $JHKL'$ color-color diagrams. The infrared colors of five of the nine objects, LkH$\alpha$~234, Objects~B, C, D, and F, are plotted here. 
The three dotted lines denote the reddening law (\cite{Koorn1983}, \cite{Bessel1988}, \cite{Martin1990}). If the color of an object is more red than that expected by interstellar extinction (the dotted lines), the object must have infrared excess in addition to its photospheric emission. Class~I sources, which are in the stages of an accreting protostar (e.g., \cite{Lada1987}), are usually located to the right of the rightmost dotted line. Class~II sources, which are objects like classical T Tauri stars with optically thick disks (e.g., \cite{Lada1987}), exist in the region between the rightmost and middle dotted lines. $L'$ band observations are useful for characterizing the infrared excesses of YSOs because deeply embedded objects can be more easily classified by their $K-L'$ colors than their $H-K$ colors (\cite{Lada2000}). The $JHKL'$ color-color diagram [Figure \ref{fig:fig_7} (b)] indicates that LkH$\alpha$~234 is more red than other objects (Objects~B, D, and F), and it is located in the region where many Class~I stars are located.
For Object~D, the lower limit of the $J-H$ color is shown in Figure \ref{fig:fig_7}. The typical visual extinction toward the LkH$\alpha$~234 region is $A_{V}=$~3.4 (\cite{Hillenbrand1992}). If Object~D is a main-sequence star, its $J-H$ color should be around 3--4 magnitudes and its visual extinction should be $\sim$20 magnitudes based on the $JHKL'$ color-color diagram. Because Object~D is near the wall of the cavity of the reflection nebula, its high extinction may be because it is located behind the nebula. Therefore, Object~D may be a YSO, but we cannot exclude the possibility that it is a main-sequence star. From the color-color diagrams, Objects~B, C, and F are either main-sequence stars or Class~III stars, which are objects like weak-lined T Tauri stars (e.g., \cite{Lada1987}). However, the fluxes of Objects~B and C are significantly different than those from previous observations. We will discuss these objects in the next section. 

The stars we detect appear to be in distinct evolutionary phases from Class~I to Class~III or main-sequence, although they are located within the small region around LkH$\alpha$~234. A similar characteristic has been observed in other star-forming regions. For example, \citet{Padgett2004} found that a region of 0.7~pc$^{2}$ in L1228 contains a total of five sources in various evolutionary stages. Note that the formation of NW1 and NW2 has occurred near LkH$\alpha$~234, which is a relatively massive B5--7 type star, although the real separation between them remains unknown. 

\subsubsection{Variable Sources}

Objects~B and C became brighter than the previous observations by \citet{Wein1994} in the $H$ and $K$ bands. The magnitudes of Object~B decreased by 0.67 $\pm$ 0.24 magnitudes in the $H$ band and 0.51~$\pm$~0.16 magnitudes in the $K$ band; those of Object~C decreased by 1.83~$\pm$~0.23 magnitudes in the $H$ band and 0.98~$\pm $~0.16 magnitudes in the $K$ band. Therefore, Object~B and C are considered to be variable like typical YSOs. 

We now discuss the possible causes of this magnitude variation. First, we consider the temporal variation of dust extinction. If the amount of the circumstellar and interstellar dust along the line of sight decreased after 1994, the object would become brighter. Adopting $A_H$~$=$~0.175 and $A_K$~$=$~0.112 when $A_V=1$ (\cite{Rieke1985}), the increment of the $H$ magnitudes $\Delta m_H= +$0.67 for Object~B suggests that $A_V$ changed by $\Delta A_V=$ $+$3.82, corresponding to an increase in the $K$-band magnitude of $\Delta m_K= +$ 0.43, which is in good agreement with the observations. In contrast, $\Delta m_H=$~$+$1.83 for Object~C and the same calculation suggests $\Delta A_V= +$~10.46 corresponding to $\Delta m_K$ $\sim +$1.17, which is not consistent with the observations. Because the color-color diagrams indicate that Object~B can be classified into Class~II or Class~III, there may be a circumstellar disk around Object~B. Therefore, the variability of Object~B may result from its obscuration by the inhomogeneous dust distribution in its disk. For Object~C, an infrared excess has not been detected. However, the $JHK$ color-color diagram alone cannot exclude the possibility of the existence of a disk (\cite{Lada1992}). 

Next, we consider the possibility that the variation occurs by the characteristics of the stellar surface. Cool spots are frequently observed for weak-line T Tauri stars, dwarfs, and the Sun. The temperatures of such cool spots are lower than the effective temperatures of their stars by $\sim$1000~K (\cite{Stahler2004}), and these magnitude variations are caused by stellar rotation. To estimate the amplitude of magnitude variation for the case when a cool spot exists on the stellar surface, we used an equation from \citet{Stahler2004}: $\Delta m_{\lambda }= -2.5~$log$[1-f{_B}[1-B_{\lambda }(T_{spot})/B{_\lambda }(T_{star})]]$. The area ratio of the cool spot can be assumed to be 20--40 percent (${f}_B$ $=$0.2--0.4) of the stellar surface (\cite{Stahler2004}, \cite{Hatzes1995}, \cite{Vrba1988}). In this case, the variation of magnitudes $\Delta m$ does not strongly depend on $f_B$. We derive the effective temperatures of Objects~B and C from the $J$ magnitudes by using the model of Baraffe et al. (\yearcite{Baraffe1998}; \yearcite{Baraffe2001}). The age was assumed to be 10$^6$~yr. Although the estimated age of LkH$\alpha$~234 is $\sim$10$^5$~yr (\cite{Hillenbrand1992}), Objects~B and C may not be so young based on the color-color diagrams. If their age is $10^6$~yr, their spectral types are F2--G2 (Baraffe et al. \yearcite{Baraffe1998}; \yearcite{Baraffe2001}). When these spectral types are assumed, the amplitudes of variation are $\sim$0.25~magnitudes in both the $H$ and $K$ bands. This variation is too small to account for the observed magnitude variations in Objects~B and C; therefore, the cool spots are probably not responsible for these variations.

Finally, we consider the possibility that hot spots cause magnitude variability. Hot spots are typically seen in classical T~Tauri stars, and they are produced by the mass accretion from a circumstellar disk onto the stellar surface. The typical temperature of hot spots is $\sim$7000~K if the stars are T Tauri stars, and hot spots cover on the order of 1~percent of the stellar surface area (\cite{Stahler2004}). We calculated the variation by using an equation from \citet{Stahler2004}: $\Delta m_{\lambda } {\sim}-2.5~($log~$e) [(hc/\lambda k_{B})(1/T_{star}-1/T_{spot})]$. For the case that the spectral types are F2--G2 and the temperatures of the hot spots are $\sim$7000~K, the magnitude variations in the $H$ band are $\Delta m_H=$~1.5~$\pm$~0.15 and in the $K$ band are $\Delta m_K=$~1.0 $\pm$ 0.1. This estimate demonstrates that hot spots can cause larger magnitude variations than cool spots. Therefore, the variation seen in Object~C may be caused by a hot spot on the stellar surface and not by circumstellar dust extinction because the hot spot model better reproduces the large amplitude of magnitude variation. Observations of periodicity will be required to discriminate between these two possibilities. In summary, the variation of Object~B is likely caused by dust extinction, whereas that of Object~C is likely caused by either dust extinction or hot spots.

\subsection{Jet-like Structures}

\begin{figure}
  \begin{center}
    \FigureFile(170mm,170mm){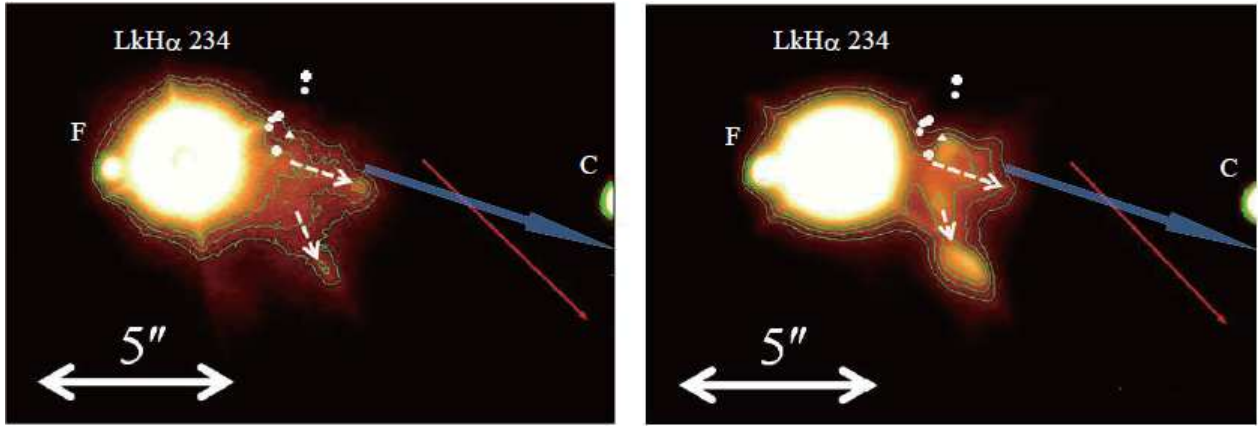}
  \end{center}
  \caption{(a) $J$-band and (b) $H$-band images around LkH$\alpha$~234. North is up and east is to the left. The white dashed arrows indicate two jet-like structures located west of LkH$\alpha$~234. The PAs of these structures are \timeform{251D-255D} and \timeform{214D-217D}. The contour levels of these $J$ and $H$ images are 64~$\sigma$ to 128~$\sigma$ and 51.5~$\sigma$ to 110~$\sigma$, respectively. Blue and red arrows represent [S~II] and H$_2$ jets, respectively (\cite{Ray1990}, \cite{Cabrit1997}). 
Crosses indicate Object~G, NW1 and NW2. Filled circles indicate VLA~2, VLA~3A and VLA~3B. 
A filled triangle indicates the mid-infrared object IRS~6 discovered by \citet{Cabrit1997}. The two jet-like structures point to neither LkH$\alpha$~234 nor NW1 but to Object~G. The position of the [S~II]~jet is roughly overlapped with those of the jet-like structures and H$_2$ jet, although the [S~II] jet is very diffuse. The H$_2$~jet does not point to LkH$\alpha$~234, Object~G, or NW1, but to NW2. }\label{fig:fig_8}
\end{figure}

As shown in Figure \ref{fig:fig_8}, two jet-like structures are seen in the $J$ and $H$-band images. These jet-like structures are located west of LkH$\alpha$~234 with PAs of  \timeform{253D} $\pm$~\timeform{2D} and \timeform{215D} $\pm$~\timeform{2D}. Because the [Fe~II] emission line may contaminate the fluxes in the $J$ and $H$ bands, these structures may represent real jets such as those observed in L~1551 IRS~5 (\cite{Itoh2000}, \cite{Pyo2002}). These two jet-like structures point neither to LkH$\alpha$~234 nor NW1 but to Object~G. The position of the northern jet-like structure is consistent with that of a [S~II] jet detected by \citet{Ray1990}, although the [S~II] jet is more diffuse in their images perhaps partly due to lower spatial resolution caused by the seeing of $1\farcs5$. [S~II] and [Fe~II] emissions usually have a similar molophology, and the emission is more compact in [Fe~II] than in [S~II] (e.g., \cite{Lopez2010}). The PA of the [S~II] jet was \timeform{252D}, which is similar to that of our jet-like structure (PA $=$ \timeform{253D} $\pm$~\timeform{2D}), but it is not clear whether this [S~II] jet originates from Object~G or LkH$\alpha$~234 because of the lower resolution. A H${}_2$ jet at PA $=$~\timeform{226D} was observed under the $0\farcs5$ seeing by \citet{Cabrit1997}, and the position is roughly overlapped with that of the diffuse [S~II] jet described above. However, the PA is different, and the H$_2$ jet does not point to LkH$\alpha$~234, Object~G, or NW1, but to NW2. It is difficult to infer the driving source from the PA alone. Note that H$_2$ and [Fe~II] jets show different spatial distributions (e.g., \cite{Lopez2010}), and jets sometimes bend and do not direct to the parent stars (e.g., \cite{Itoh2000}). High-spatial resolution spectroscopy will help identify the driving sources of the jets (e.g., \cite{Pyo2002}).




\subsection{Two Candidates for the Star Illuminating the Reflection Nebula}

A remarkable reflection nebula is seen west of LkH$\alpha$~234. We consider two candidates besides LkH$\alpha$~234 for the star illuminating the nebula: Object~G and NW1.

The position of Object~G implies that it is the source of an outflow. First, it almost corresponds to the central position of the 2~$\micron$ polarization determined by \citet{Wein1994} (Table \ref{tab:table_3}). In addition, the jet-like structures detected in the $J$ and $H$ bands point toward Object~G. However, we are unable to exclude the possibility that Object~G is a background star because it is difficult to estimate its luminosity using only its $H$- and $K$-band magnitudes. Because the magnitudes in these bands may include the infrared excess, it is difficult to derive the amount of extinction, and thus its spectral type and distance. Therefore, it is also unclear if Object~G is luminous enough to illuminate the reflection nebula in this region. Note, however, because Object~G may be located near the mid-plane in the disk of NW1 or Object~G itself, the extinction toward Object~G may be quite high if it resides behind the disk mid-plane. If Object~G is associated with this region and affected only by the typical extinction toward this region ($A_V$ $=$3.4, \cite{Hillenbrand1992}), its mass is approximately that of a brown dwarf (with a spectral type of later than M6) based on evolutionary tracks (Baraffe et al. \yearcite{Baraffe1998}; \yearcite{Baraffe2001}, \cite{Siess2000}). In this case, Object~G might be the lowest-mass object found in this region. Previous studies have confirmed the existence of jets ($\sim$100~AU) and small-scale ($\sim$1000~AU) outflows driven by brown dwarfs and very low-mass stars (\cite{Phan-Bao2008}, \cite{Lee2009}, \cite{Whelan2009}). The jet-like structures and monopolar nebula we detected have larger sizes ($>$~1000~AU); therefore, if Object~G is a brown dwarf, it might have difficulty driving the observed jets and outflow. In contrast, one might think that a higher extinction toward Object~G is plausible as a high extinction of $A_{V}=15$ is derived for Object~F from the color-color diagrams (Figure \ref{fig:fig_7}). Assuming the same $A_{V}$ as Object~F, the spectral type of Object~G is estimated to be M5 or K3 depending on the evolutionary tracks (Baraffe et al. \yearcite{Baraffe1998}; \yearcite{Baraffe2001}, \cite{Siess2000}). Therefore, we cannot exclude the possibility that Object~G ejects these jets and outflow. 

\begin{figure}
  \begin{center}
    \FigureFile(160mm,160mm){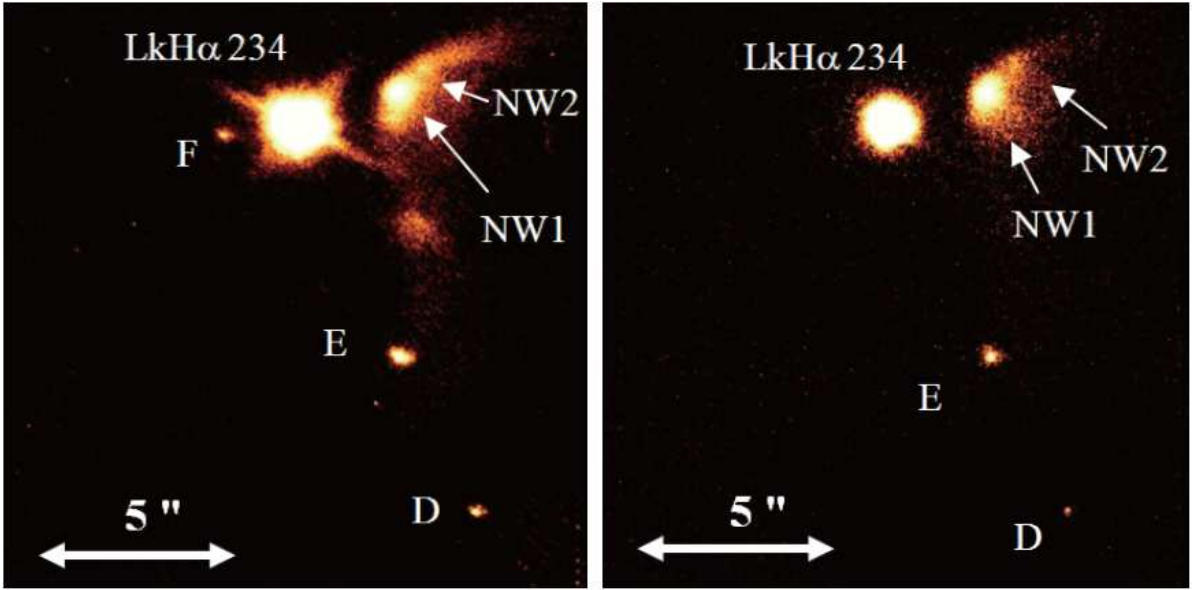}
  \end{center}
  \caption{(a) $L'$-band image (left panel) and (b) $M'$-band image (right panel) of the LkH$\alpha$~234 region. These images are expressed in logarithmic scale. North is up and east is to the left. Five objects (Objects~D, E, F, NW1, and NW2) are detected around LkH$\alpha$~234. NW1 is bright in these bands. The display level of (a) is 1~$\sigma$ to 8.5~$\sigma$, while that of (b) is 1~$\sigma$ to 5.7~$\sigma$. }\label{fig:fig_2}
\end{figure}

\begin{figure}
  \begin{center}
    \FigureFile(150mm,120mm){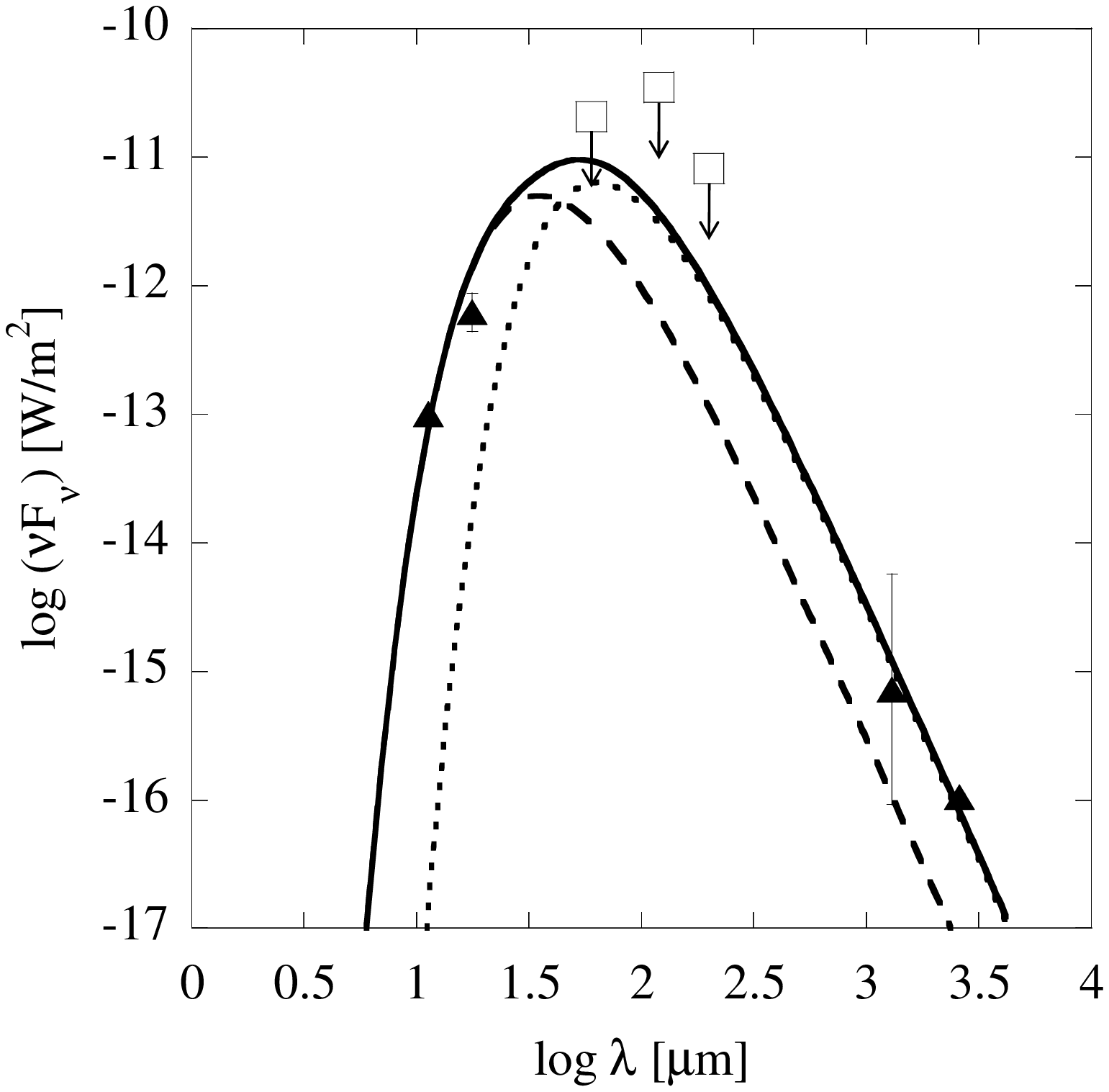}
  \end{center}
  \caption{Spectral energy distribution of NW1. The mid-infrared (11.8~$\micron$ and 17.65~$\micron$) data are from the present study. The far-infrared (60, 120 and 200~$\micron$) and millimeter (1.3 and 2.6~mm) data are from \citet{Abraham2000} and \citet{Fuente2001}, respectively. The far-infrared observations are only upper limits because NW1 is unresolved. The dotted curve represents a gray-body sphere with temperature of 52~K and an outer radius of 667~AU; this curve is fitted to the millimeter data observations (cold dust components). The dashed curve represents warm dust components (at 100~K and 147~AU). The solid curve shows the result of adding the cold and warm dust components. These fittings were adopted using $\beta =1$ and $\tau_{100} = 1$ (\cite{O$'$linger2006} and \cite{Tachihara2007}). }\label{fig:fig_9}
\end{figure}

NW1 is bright in the $L'$, $M'$, and mid-infrared bands (Figures \ref{fig:fig_2} and \ref{fig:fig_5}), consistent with the conclusion by \citet{Fuente2001} that NW1 is an embedded young star. Figure \ref{fig:fig_9} shows the spectral energy distribution (SED) of NW1; it was constructed using our near- and mid-infrared photometry and the millimeter flux densities at 1.3 and 2.6~mm obtained with the IRAM 30~m telescope (\cite{Fuente2001}). Far-infrared data (60, 120, and 200~$\micron$) taken with the Infrared Space Observatory's (ISO) photometer (ISOPHOT) were also included in the SED (\cite{Abraham2000}), but they are considered to be upper limits because ISO's large beam could not resolve NW~1 and NW~2. The dotted curve shows a gray-body sphere with a temperature of 52~$K$ and an outer radius of 667~AU. We adopted $\beta$ = 1 and $\tau_{100}$ = 1 for $\tau_{\nu }$= $\tau_{100}^{-\beta }$ where $\tau_{100}$ is $\tau$ at $\lambda = 100~\micron$ (\cite{O$'$linger2006}, \cite{Tachihara2007}). This curve is fitted to the millimeter data points, which represent the cold dust component (\cite{Fuente2001}). The dashed curve shows the warm dust component with 100~K and 147~AU for the temperature and outer radius, respectively. Finally, the solid curve shows a sum of the cold and warm dust components.

SED of NW1 that includes our new observations also indicates that NW1 is an embedded YSO. The mid-infrared and millimeter data can be fitted by combining the cold and warm dust components. The radius of the sphere derived from the fitting (667~AU) is consistent with the millimeter interferometric observations when the area of the continuum emission from cold dust grains is about 1000~AU in radius. In contrast, our near-infrared ($L'$ and $M'$~bands) and mid-infrared (11.8 and 17.65~$\micron$) images show that the radius (FWHM) of NW1 is $\sim$260~AU and $\sim$140~AU, respectively. NW1 is extended in the $L'$ and $M'$~bands, but scattered light may contribute to this; its MIR size better represents the thermal emission from warm dust. Therefore, the radius of the warm dust component inferred by the SED fitting (147~AU) is consistent with that infferred from our observed mid-infrared size ($\sim$140~AU). From the dual-graybody fitting (solid curve), NW1's luminosity is estimated to be $\sim$700~$L_{\odot}$. The spectral type corresponding to this luminosity is B6--7 assuming it is a main-sequence star, which is similar to the B5--8 spectral type ($\sim$500~$L_{\odot}$) suggested by \citet{Fuente2001}.     

The following objects were also observed around LkH$\alpha$~234: a redshifted CO outflow, a blueshifted [S~II] jet, four radio continuum sources (VLA~1, VLA~2, VLA~3A, and VLA~3B), and associated H${}_2$O maser spots (\cite{Trini2004}, \cite{Fuente2001}, \cite{Ume2002}; Table \ref{tab:table_3}). H${}_2$O masers are usually found around accretion disks and outflows. As shown in Tables \ref{tab:table_1} and \ref{tab:table_3}, the position of VLA~2 agrees well with that of NW2. NW2 is more faint in the near- and mid-infrared, implying that it is more embedded than NW1, or that it simply has a lower luminosity regardless of its evolutionary phase. \citet{Trini2004} favored the interpretation that VLA~2 is driving a large-scale CO and [S~II] outflow based on the distribution of maser spots around it; however, the [S~II] jet does not exactly point toward NW2 as seen in Figure \ref{fig:fig_8}. In contrast, NW1 is another candidate for a driving source of the CO outflow, as \citet{Fuente2001} originally suggested based on their interferometric CO $J=1 \rightarrow 0$ images. VLA~3A and VLA~3B are located near the position of NW1. They are thought to form a close ($\sim$220~AU) binary system, and they seem to drive a thermal radio jet (\cite{Trini2004}). NW1 appears as one source in our images, and its position is more consistent with that for VLA~3B. Moreover, NW1 is also located near the central position of the 2~$\micron$ polarization (\cite{Wein1994}; Table \ref{tab:table_3}), and the morphology of the reflection nebula in the near-infrared strongly suggests that NW1 is associated with the monopolar nebula (Figures \ref{fig:fig_1} and \ref{fig:fig_2}). Consequently, it is possible that NW1 is a YSO of spectral type B6--7 that actively causes the large-scale CO outflow. If so, it could illuminate the reflection nebula in this region.

In any case, LkH$\alpha$~234 is not considered to drive the large-scale CO outflow and various jets (the [S~II] and H$_2$ jets and the possible jets detected at $J$ and $H$) observed in this region. We suggest that the source illuminating the reflection nebula is NW1, but we cannot completely exclude the possibility that it is Object~G. 

\subsection{Companion Candidate for LkH$\alpha$~234}

Object~F was identified for the first time. This object is located $1\farcs9$ ($\sim$2400~AU) from LkH$\alpha$~234, and it is likely a star without infrared excess (see subsection 4.2). Assuming that the extinction $A_V$ is 15 magnitudes (from the color-color diagrams of Figure \ref{fig:fig_7}) and the age is 10$^{6}$~yr, we attempted to estimate the mass of Object~F from its $J$-band magnitudes using model isochrones (\cite{Palla1999}, \cite{Siess2000}). We assumed an age of 10$^6$~yr for Object~F since the age of LkH$\alpha$~234 (10$^5$~yr) seems too young for the source without excess near-infrared emission. The estimate suggests that the spectral type of Object~F is A2--A5 (2.0--2.5~${M}_{\odot}$) if Palla \& Stahler's (\yearcite{Palla1999}) method is adopted, whereas it is $\sim$K4 if Siess et al's (\yearcite{Siess2000}) method is adopted. Therefore, Object~F may be a low- or intermediate-mass star associated with this star-forming region, and a companion candidate for LkH$\alpha$~234 separated from it by 2400~AU. Herbig Ae/Be systems are frequently (30--75 percent) found in binary systems with separations between 30 and 4000~AU (\cite{Wheelwright2010}, \cite{Carmona2007}, \cite{Baines2006}, \cite{Leinert1997}). The mass ratio of LkH$\alpha$~234 to Object~F is 0.08--0.29, consistent with the results of previous observations of Herbig Ae/Be stars. If Object~F is a companion, judging from the separation, it might have formed via disc fragmentation, which can produce a binary star with a higher mass ratio than that produced via core fragmentation (\cite{Wheelwright2010}, \cite{Kouwenhoven2009}).

The offset distance of Object~G from LkH$\alpha$~234, which is $2\farcs3$ ($\sim$2900~AU), is similar to that of Object~F. If Object~G is a companion of LkH$\alpha$~234 and its mass is the same as that of a brown dwarf (Section 5.1), the mass ratio is below 0.01. This mass ratio is small compared with previous observations. In contrast, if the spectral type of Object~G is $\sim$M5 or K3 (Section 5.1), the mass ratio is 0.03--0.1, which is slightly smaller than the mass ratio of Object~F. There is little understanding about the complete census on the mass ratio and the separation because higher contrast is required to investigate fainter and closer companions around the Herbig Ae/Be stars (\cite{Wheelwright2010}, \cite{Baines2006}). In the case of Object~G, it would be hard to determine if it is physically bound to LkH$\alpha$~234, and to estimate the correct mass; however, our result demonstrates that AO imaging is useful for increasing the lower-mass sample of companion stars to Herbig Ae/Be stars to constrain the binary formation mechanism.


\section{Summary}
We observed the Herbig Ae/Be star LkH$\alpha$~234 at near-infrared wavelengths ($J$, $H$, $K$, $L'$ and $M'$ bands) using CIAO on the Subaru Telescope. We obtained a high-angular resolution (0$\farcs$2) with AO. The images reveal the complex morphology of the reflection nebula at approximately 3$\arcsec$ from LkH$\alpha$~234 that extends more than 10$\arcsec$ westward. In addition, eight (Objects~B, C, D, E, F, G, NW1, and NW2) YSO candidates were detected 2$\arcsec$--11$\arcsec$ from LkH$\alpha$~234, and four (Objects~E, F, G, and NW2) of them were identified for the first time in the near-infrared. Object~D is a Class II source or a main-sequence star, based on the $JHK$ and $JHKL'$ color-color diagrams. Objects~B and C are likely variable stars based on a comparison with previous photometry of them, and they could be YSOs. The variation of Object~B is explained by variable extinction by circumstellar dust, and that of Object~C may be caused by dust extinction or hot spots. LkH$\alpha$~234, NW1, and NW2 were also detected in the 11.8~$\micron$ (SiC) and 17.65~$\micron$ images obtained with the LWS on the Keck Telescope. NW1 and NW2 are very bright in the mid-infrared bands, but they could not be seen in the $J$, $H$, and $K$-band images. Therefore, they are likely embedded young stars. We found that these stars are in various evolutionary phases from Class~I to Class~III or main-sequence, and they are located within a small (0.14~pc$^2$) region around LkH$\alpha$~234. Furthermore, we suggest that the source illuminating the reflection nebula is NW1 and not LkH$\alpha$~234 although Object~G may be another candidate. NW1 is located near the central position of the 2~$\micron$ polarization and it is near the origin of the monopolar nebula. There is a possibility that NW1 is a YSO whose spectral type is B6--7 based on its SED and that it illuminates the reflection nebula in this region. Object~F may be a companion star of LkH$\alpha$~234 with a separation of $1\farcs9$ ($\sim$2400~AU) and a low-mass star with a spectral type of A2--A5 or $\sim$K4 depending on the model isochrones. The mass ratio of LkH$\alpha$~234 to Object~F is then 0.08--0.29, consistent with previous observations of Herbig Ae/Be binaries. Moreover, Object~G has a similar separation from LkH$\alpha$~234 ($\sim$2900~AU), and if it is a companion, the mass contrast may be even larger.

We thank D. Kontopoulos for the careful reading of our manuscript. We are also grateful to T. Kohyama for the kind support.

\newpage

\bigskip



\clearpage

\clearpage

\begin{longtable}{llllllll}
  \caption{Summary of near-infrared observations}\label{tab:table_log}
  \hline              
  Date (yy--mm--dd) & Band & $T_{exp}$ [s] & Coadd & $T_{tot}$ [s] & Mask size [$\arcsec$] & Standard star & Seeing [$\arcsec$]  \\ 
\endfirsthead
  \hline
  Date (yy--mm--dd) & Band & $T_{exp}$ [s] & Coadd & $T_{tot}$ [s] & Mask size [$\arcsec$] & Standard star & Seeing [$\arcsec$] \\  
\endhead
  \hline
\endfoot
  \hline
\multicolumn{1}{@{}l}{\rlap{\parbox[t]{.90\textwidth}{\small
  Columns~1 and 2 list the dates (year-month-date) and the observational bands. Columns~3 and 4 represent the exposure time and the number of coadds for each frame. $T_{tot}$ in Column~5 gives the total exposure time. Column~6 gives the diameter of the occulting mask; we utilized a circular Lyot stop at the pupil plane to block out 20~percent of the pupil diameter. Column~7 lists the photometric standard stars (\cite{Leggett2003}, \cite{Leggett2006}, and \cite{Cohen1999}). Column~8 gives the natural seeing in optical. }}} 
\endlastfoot
  \hline
 06-07-15 & $J$ & 10 & 1 & 60 & 0.8 & FS~150 & 0.8  \\
   & $H$ & 6 & 1 & 300 & 0.8 & FS~150 & 0.8  \\
   & $K$ & 6 & 1 & 300 & 0.8 & FS~150 & 0.8  \\ 
 08-05-25 & $J$ & 10 & 1 & 300 & 0.8 & FS~150 & 0.5  \\
 08-07-10 & $L'$ & 0.18 & 10 & 252 & -- & HD129655 & 1.3  \\
   & $M'$ & 0.18 & 10 & 468 & -- & HD129655 & 1.3  \\   
 04-07-26 & 11.8 & 0.01 & 28500 & 285 & -- & HD197989 & -- \\ 
 04-08-28 & 17.65 & 0.01 & 28500 & 285 & -- & HD197989 & -- \\ 
\end{longtable}



{\setlength{\tabcolsep}{0.5pt}\small
\begin{longtable}{llllllllll} 
  \caption{Positions and magnitudes of detected objects.}\label{tab:table_1}
  \hline              
  Object & Dis. [$\arcsec$] & P.A. [\timeform{D}] & ~~~~ $J$ & ~~~~$H$ &~~~~ $K$ & ~~~~$L'$ & ~~~~$M'$ & ~~~SiC [Jy] & 17.65~$\micron$ [Jy] \\ 
\endfirsthead
  \hline
  Object & Dis. [$\arcsec$] & P.A. [\timeform{D}] & ~~~~$J$  & ~~~~$H$ & ~~~~$K$ & ~~~~$L'$ & ~~~~$M'$ & ~~~SiC [Jy] & 17.65~$\micron$ [Jy] \\  
\endhead
  \hline
\endfoot
  \hline
\multicolumn{1}{@{}l}{\rlap{\parbox[t]{.95\textwidth}{\small
  The distances and PAs are from LkH$\alpha$~234. The uncertainty in the offset distance and in the position angle at the near-infrared wavelengths are $0\farcs05$ and \timeform{1.7D}, respectively. Those at the mid-infrared wavelengths are $0\farcs15$ and \timeform{2.6D}, respectively. The figures in parentheses are the uncertainties ($\pm$) in the photometric values. The detection limits in the $J$, $H$, $K$, $L'$, and $M'$ bands are 18.5~$\pm$~0.2 magnitudes, 19.0~$\pm$~0.3~magnitudes, 17.7$\pm$~0.2~magnitudes, 12.1~$\pm$~0.2 magnitudes, and 10.0~$\pm$~0.2 magnitudes, respectively. The magnitudes of LkH$\alpha$~234 in the $JHK$ bands are from the 2MASS All-Sky Catalog. }}} 
\endlastfoot
  \hline
  LkH$\alpha$~234 & ~~0.0 & ~~0.0 & 9.528 (0.02) & 8.201 (0.02) & 7.081 (0.02) & 5.23 (0.05) & 4.83 (0.18) & 3.53 (0.15) & 4.95 (0.50)   \\
  B & 8.28 & 11.16 & 15.05 (0.06) & 13.43 (0.04) & 12.39 (0.04) & 11.45 (0.06) & 11.19 (0.36) & ~~~~-- & ~~~~--  \\
  C & 11.12 & 265.3 & 13.15 (0.06) & 12.25 (0.03) & 11.88 (0.04) & ~~~~-- & ~~~~-- & ~~~~-- & ~~~~--  \\
  D & 10.70 & 205.4 & $>$18.5 & 15.50 (0.10) & 13.17 (0.06) & 11.40 (0.05) & 11.10 (0.56) & ~~~~-- & ~~~~-- \\
  E & 6.37 & 204.4 & $>$18.5  & $>$19.0 & 15.04 (0.30) & 10.39 (0.05) & 9.40 (0.22) & ~~~~-- & ~~~~--  \\
  F & 1.89 & 96.9 & 15.45 (0.25) & 13.29 (0.23) & 11.99 (0.22) & 11.6 (0.06) & $>$11.37 & ~~~~-- & ~~~~--  \\
  G & 2.28 & 275.9 & $>$18.5 & 18.26 (0.40) & 16.46 (0.20) & ~~~~-- & ~~~~-- & ~~~~-- & ~~~~--  \\
  NW1 & 2.52 & 295.7 & ~~~~-- & ~~~~-- & ~~~~-- & 9.50 (0.06) & 8.32 (0.2) & 0.38 (0.02) & 3.64 (0.40)  \\
  NW2 & 3.69 & 305.5 & ~~~~-- & ~~~~-- & ~~~~-- & 10.58 (0.06) & 9.56 (0.4) & 0.03 (0.01) & 0.95 (0.30) \\
\end{longtable}
}

\begin{longtable}{ll}
  \caption{Corresponding objects in previous studies.}\label{tab:table_4}
  \hline              
  Near/Mid IR objects (in this paper) & Corresponding object \\ 
\endfirsthead
  \hline
  Near/Mid IR objects & Corresponding object  \\  
\endhead
  \hline
\endfoot
  \hline
  \multicolumn{1}{@{}l}{\rlap{\parbox[t]{.70\textwidth}{\small}}} 
\endlastfoot
  \hline
  Object~B & IRS~2 (e.g., \cite{Wein1994}) \\
  Object~B & IRS~4 (e.g., \cite{Wein1994}) \\
  Object~E & C3 (\cite{Tofani1995}) \\
  LkH$\alpha$~234 NW1 & PS1 (\cite{Wein1994}) \\
                      & 10~\micron ~companion or IRS~6 \\
                      & (\cite{Cabrit1997}, \cite{Fuente2001})\\
                      & VLA~3A+B (\cite{Trini2004}) \\
                      & LkH$\alpha$~234 NW (\cite{Polomski2002}) \\
  LkH$\alpha$~234 NW2 & VLA~2 (Tofani1995) \\
\end{longtable}



\begin{longtable}{llll}
  \caption{Positions of outflow source candidates from other observations.}\label{tab:table_3}
  \hline              
  Object & Distance [$\arcsec$] & P.A. [\timeform{D}] & Author \\ 
\endfirsthead
  \hline
 Object & Distance & P.A. & Author  \\  
\endhead
  \hline
\endfoot
  \hline
  \multicolumn{1}{@{}l}{\rlap{\parbox[t]{.65\textwidth}{\small
  Offset distances and offset position angles are from LkH$\alpha$~234.}}} 
\endlastfoot
  \hline
  PS 1 & 3 & 290-340 & \cite{Wein1994} \\
  PS 1 & 2.58 & -- & \cite{Wein1996} \\
  10 $\micron$ companion & 2.7 & 280-286 & \cite{Cabrit1997} \\
  VLA 1 & 5.91 & 312 & \cite{Tofani1995} \\
  VLA 2 & 3.50 & 300 & \cite{Tofani1995} \\
  VLA 3A & 2.23 & 292 & \cite{Trini2004} \\
  VLA 3B & 2.40 & 295 & \cite{Trini2004} \\
\end{longtable}



\end{document}